\documentclass[12pt]{iopart}

\usepackage[T1]{fontenc}
\usepackage[utf8x]{inputenc}
\usepackage[english]{babel}
\usepackage{graphicx}
\usepackage{comment}
\usepackage{xcolor}

\expandafter\let\csname equation*\endcsname=\relax 
\expandafter\let\csname endequation*\endcsname=\relax 
\usepackage{amsmath,amssymb,amsthm, amsfonts} 

\newcommand{\bx}{\mathbf{x}}

\newcommand{\by}{\mathbf{y}}
\newcommand{\bxi}{\mathbf{\xi}}

\begin{document}

\selectlanguage{english}

\title{Sparse Representations, Inference and Learning}

\author{
C Lauditi$^1$, E Troiani$^{2,}$ and M Mézard$^3$}

\address{$^1$ Department of Applied Science and Technology, Politecnico di Torino, Italy}
\address{$^2$ Statistical Physics of Computation Laboratory, EPFL, Switzerland}
\address{$^3$ Department of Computing Sciences, Bocconi University, Milano}

\ead{emanuele.troiani@epfl.ch}

\begin{abstract}

In recent years statistical physics has proven to be a valuable tool to probe into large dimensional inference problems such as the ones occurring in machine learning.  Statistical physics provides analytical tools to study fundamental limitations in their solutions and proposes algorithms to solve individual instances. 
In these notes,
based on the lectures by Marc Mézard in 2022 at the summer school in Les Houches, 
we will present a general framework that can be used in a large variety of problems with weak long-range interactions, including the compressed sensing problem, or the problem of learning in a perceptron. We shall see how these problems can be studied at the replica symmetric level, using developments of the cavity methods, both as a theoretical tool and as an algorithm.
\end{abstract}

\section{Introduction}
Equilibrium statistical physics that explores high-dimensional probability distributions, has significantly evolved over the past fifty years due to advances in the theory of disordered systems. This has led to its application across a wide spectrum of problems. Our discussion will initially focus on two illustrative examples: one from the field of machine learning and the other from information processing, both of which exemplify this category. Subsequently, we will describe a broader class of problems that not only incorporate the above examples, but also a variety of other scenarios involving numerous variables engaging in weak long-range interactions. We will systematically detail how this inclusive set of problems can be examined at replica symmetric level. This examination employs the cavity method, alternatively known as belief propagation (BP), which gives rise to a suite of algorithms referred to as approximate message passing (AMP). We will provide an in-depth discussion on how numerous problems, which typically require an exponential computational expense relative to the number of variables, can be addressed more efficiently with polynomial complexity using these techniques.

\subsection{An Example from Machine Learning: the generalized Perceptron in ridge regression setting}

How can one address some issues of machine learning problems using techniques inspired by physics? We first give a simple example of that in an emblematic setting: the generalized perceptron solving a ridge regression task. The problem can be summarised as follows: assume you have a dataset $\mathcal{D}$ of $P$ samples $\mathcal{D} = \{y^\mu, \mathbf{F}^\mu\}_{\mu = 1,...,P}$, being $\mathbf{F}^\mu$ the $N$-dimensional data while $y^\mu$ the set of labels for these. The generative recipe according to which data are labelled can be enforced by introducing some ``teacher'' structure. In practice, for each data point 
    $\mathbf{F}^\mu$, the teacher generates the label according to the rule
\begin{equation} \label{eq:regression_process}
    y^\mu = \phi\left(\mathbf{F}^\mu \cdot \mathbf{x}^*\right) + \eta^\mu
\end{equation}
where $\phi (\cdot)$ is called activation function while $\mathbf{x}^*$ are the parameters that characterize the teacher. We have also incorporated some random Gaussian noise in the expression by adding $\eta^\mu \sim \mathcal{N}(0,\Delta^2)$, which models the situation in which it is possible for the student to infer the teacher rule even if it is corrupted independently of the examples. A simple way to rephrase the problem is that the student network with parameters $\mathbf{x}$ would like to recover the parameters $\mathbf{x}^*$ of the teacher by minimising an error function $E(\mathbf{x},\mathcal{D})$
\begin{equation}\label{eq::E_perceptron}
    \min_{\mathbf{x}} E(\mathbf{x},\mathcal{D}) = \min_{\mathbf{x}}\ \left[ \sum_\mu (y^\mu - \phi(\mathbf{F}^\mu \cdot\mathbf{x}))^2 + V(\mathbf{x})\right]
\end{equation}
which measures the average squared difference between the actual and predicted values of the labels for each patterns. Precisely, in the expression~\eqref{eq::E_perceptron}, the first term pushes the student vector $\mathbf{x}$ to learn the generative process while the second one, commonly called regularisation, restricts the space of parameters that $\mathbf{x}$ can take.
The function $\phi(\cdot)$ is usually non-linear and depending on the task may reduce to commonly used versions: the $\text{sign}(x)$ in the standard perceptron case, while generalised variations can account for smoother version like $\tanh(\cdot)$ or sparsity-enhancing functions like the rectified linear unit $\text{ReLU}(x)=\max{(0,x)}$. We refer to \cite{Engel_VdB} for an introduction to perceptrons with a statistical physics perspective.

Given the prescription, we can reformulate the learning problem in a probabilistic setting as Bayesian estimation: extract  $\mathbf{F}^\mu \sim P^{\mathbf{F}}$,  $\mathbf{x}^* \sim P^{\mathbf{x}}$ and generate $\{y^\mu\}$ through the process (\ref{eq:regression_process}), then given $\mathcal{D}$ the objective is to recover $\mathbf{x}$ that minimises on average the error function $E(\boldsymbol{x},\mathcal{D})$ using the posterior probability
\begin{equation}
  P(\mathbf{x}\,|\,\mathcal{D}) = \lim_{\beta\to \infty} \frac{1}{\mathcal{Z}}\, e^{-\beta E(\mathbf{x}, \mathcal{D})}
  \label{Perceptron_Bayesian}
\end{equation}
and given some prior distribution that is factorized on the parameters $\prod_i P_{\boldsymbol{x}}(x_i)$.
Resorting to a statistical mechanics formulation of the problem, this is the well-known Boltzmann distribution and we can think of the parameters of our model as the "spin" variables while the $\beta$ parameter is the inverse of temperature. In practice, $\beta$ controls the distribution over the parameters, that in the limit $\beta\to\infty$ will concentrate on the values that minimise the error function for a fixed instance of the problem.
We will typically study the high-dimensional regime where $\mathbf{F}^\mu$ is of dimension $N \gg 1$. The thermodynamic limit is defined by taking a large number of samples $P \gg 1$, keeping the density of constraints $\alpha = P/N$ finite.

The dataset  plays the role of the quenched disorder in the system, and motivates the use of techniques from spin glass physics to deal with the problem setting, e.g. replica or cavity methods
\cite{mezard_parisi_virasoro_2004, mezard_montanari_2017}. Precisely, the typical questions in machine learning are then equivalent to questions one asks in physics: computing the optimal error is equivalent to finding the ground state energy and finding an efficient algorithm to reach the optimal configuration is equivalent to having an efficient  sampling procedure.

\subsection{An Example from Information Theory: Compressed sensing}
Compressed sensing \cite{CandesTao:05,Donoho:06} is another interesting problem in information theory and signal processing where we can apply the standard workflow of statistical physics \cite{Krzakala_2012}. It has found numerous applications, including image and video compression, medical imaging or spectral analysis and it is based on the simple idea that, through optimization, the sparsity of a signal can be exploited to recover it from far fewer samples than required by the Nyquist–Shannon sampling theorem. Indeed, in many practical applications, signals are sparse or compressible in a known basis or dictionary, meaning that they can be accurately represented with only a few non-zero coefficients.

\begin{figure}
    \centering
    \includegraphics[width=8cm]{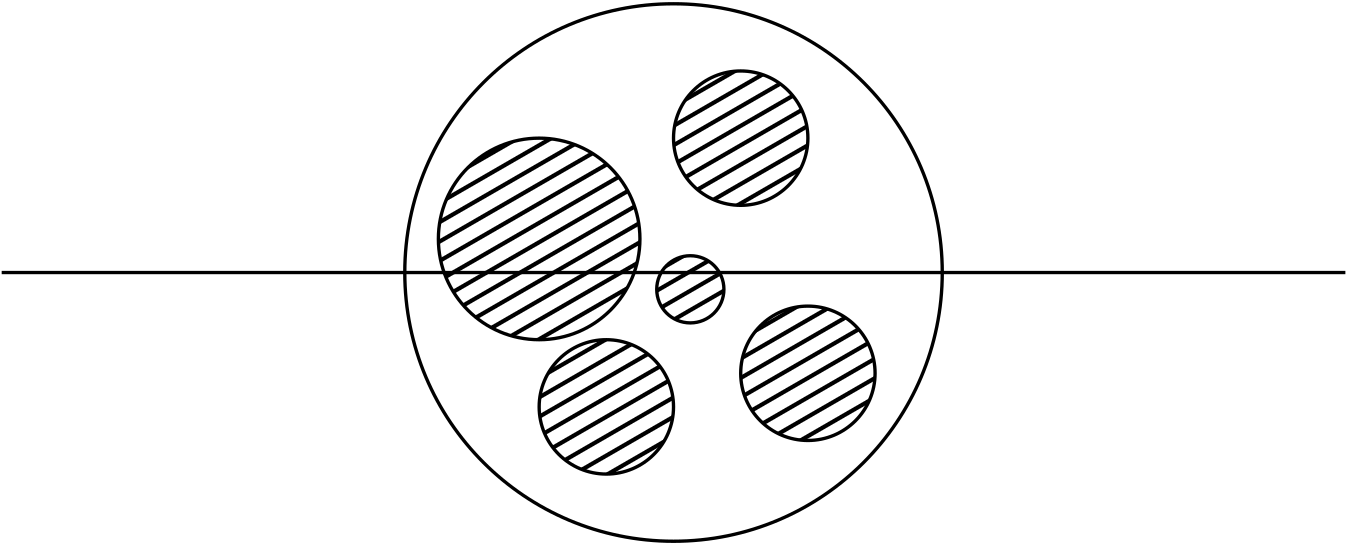}
    \caption{A binary alloy crossed by a measurement ray}
    \label{fig:alloy}
\end{figure}

With that insight in mind, let us start by a simple physical example which is a problem of tomography. Imagine you are studying a 2D binary alloy like the one in Figure \ref{fig:alloy} which would be a slice of a three-dimensional material. This is an object that is made of two materials: \textbf{A} and \textbf{B}. We can give a geometrical description of the alloy by defining it as a compact region such that to each point in the interior is given a binary value in $\{0,1\}$ that will tell whether it is made of material \textbf{A} or \textbf{B}. The task is to describe the interior of the alloy non-destructively. 

To do so we can use synchrotron light: if we shine a ray of light across the material we can measure how much the intensity decreases between the entrance and exit points of the ray in the alloy. We assume that yhe light is absorbed differently by the two materials,  each measurement will essentially give a ratio of portions of material \textbf{A} over \textbf{B} along the ray. 

The alloy we are going to study will have typical size $L$ and let us suppose our light device is able to see pixels of size $a \ll L$.
This means there will be a total of $N \propto (L/a)^2$ pixels, so the maximal entropy, defined as the number of possible alloys that our device can measure is in principle $S \propto N \log{2}$ given that the model is binary. We thus expect to fully recover our information about the alloy by doing approximately a number $N$ measurements, for example by measuring in parallel $L/a$ parallel rays, each for $L/a$ different orientations. This is the standard way of analysing a sample: we measure the absorption along $(L/a)^2$ rays, and reconstruct the composition by a Radon transform.

However, having some information on the alloy allows to devise some more clever strategies. If for example  the alloy is made of mostly material \textbf{A} with some spots of material \textbf{B} with characteristic length $\zeta$, we know that the number of possible samples is much reduced and the entropy is of order $(L/\zeta)^2$ which is much less than $(L/a)^2$. In such a case we can proceed with the measurements in a more efficient way. 

It should therefore be clear that the length $\zeta$ indicates the difficulty of our problem: if $\zeta \approx a$ the spots are too small to be seen and we are in the case without information, so the object is hard to analyse but we cannot do better than the naive procedure. Instead, if $\zeta \approx L$ the object is almost homogeneous so the problem is easy. The intermediate case
$a\ll\zeta\ll L$ is the most interesting.

This problem of tomography can be studied (see 
\cite{2013_Gouillart_etal_tomo}), but it is somewhat complicated and requires some approximations. So we will introduce here a simplified version which can be solved in all detail. This will allow us to understand the basic mechanisms of compressed sensing.
We now have $N$ variables $x_i$, generated with a factorized prior measure $P_x(\bx)=\prod_i P_x(x_i)$ which would have the role of describing our sample of alloy. As before, we do not know the variables but we perform $P$ independent measurements obtaining different results $\{y^\mu\}$ derived from the protocol
\begin{equation}
    y^\mu = \sum_{i} F_{\mu i} x_i + \eta^\mu= \mathbf{F}^\mu\cdot\bx + \eta^\mu.
\end{equation}
So, in the simplified version of the problem, we know the apparatus $\mathbf{F}^\mu = \{F_{\mu i}\}$ that is a $N$-component vector such that the measurement number $\mu$ is the projection onto $\mathbf{F}^\mu$ being $i$ the index of the pixels, and $\eta^\mu \sim \mathcal{N}(0,\Delta^2)$ the white noise corrupting the information to be recovered. Then, the probability of obtaining a measurement $y^\mu$ given a certain alloy configuration $\mathbf{x}$ is simply 
\begin{equation}
    P(y^\mu|\mathbf{x}) = \frac{1}{\sqrt{2\pi\Delta^2}}\exp{\left\{-\frac{\left(y^\mu-\mathbf{F}^\mu \cdot \mathbf{x}\right)^2}{2\Delta^2}\right\}}.
\end{equation}

In a Bayesian approach, we must combine this probability, obtained from the measurements, together with a prior that describes our a-priori knowledge (for instance, it could be a binary distribution in the binary alloy setting). Using Bayes theorem we get the posterior probability or Boltzmann measure
\begin{equation}    \label{cs-formulation}
    P(\mathbf{x}|\{y^\mu\}) = \frac{1}{\mathcal{Z}}\; P_x(\bx)\; \exp{\left\{-\frac{1}{2\Delta^2}\sum_{\mu=1}^P\left(y^\mu-\mathbf{F}^\mu \cdot \mathbf{x}\right)^2\right\}}.
\end{equation}

We have thus written the compressed sensing problem in a way that is amenable to statistical physics tools. 
Given a measurement device $\mathbf{F}^\mu$, and the results of the measurements $y^\mu$, we need to study the probability (\ref{cs-formulation}) over the possible compositions $\mathbf{x}$ of the variables. We could aim either at sampling from this probability, or at finding the most probable composition $\mathbf{x}$. Note that this is a complicated problem as we are looking at a probability in a large dimensional space. Furthermore it is a disordered system, since the probability depends on a number of  parameters  $\mathbf{F}^\mu$ and  $y^\mu$ which is of order $N$.

For practical reconstruction we need to address the algorithmic problem of studying the probability (\ref{cs-formulation}) for a given set of $\mathbf{F}^\mu$ and  $y^\mu$. On the other hand, in order to gain some insight, we might want to study analytically cases in which $\mathbf{F}^\mu$ is obtained from a random ensemble. This will allow us to study the ultimate information-theoretic reconstruction, as well as the performance of classes of algorithms, for typical cases of $\mathbf{F}^\mu$.
For instance, it is natural to consider $\mathbf{F}^\mu$ as sparse random vectors extracted independently from a distribution $P^{\mathbf{F}}$.

Let us focus  first on the noiseless $\Delta\to 0$ case, in which
\begin{equation}
 y^\mu = \mathbf{F}^\mu \cdot \mathbf{s}.
\end{equation}
In this limit the compressed sensing is simply a problem of linear algebra, i.e. from the $P$ measurements $\{y^{\mu}\}$ one wishes to recover the vector $\mathbf{s}$ by assuming the vector of measurements $\{\mathbf{F}^{\mu}\}$ to be linearly independent, which is a weak requirement when dealing with random vectors in high dimension. If $P < N$ then the problem is impossible: it is equivalent to trying to solve a system of $P$ equations in $N$ variables when it is under-determined. If $P \geq N$ the problem instead is solvable: the linear system is over-complete, so we can simply choose $N$ equations and use linear algebra techniques to find $\mathbf{x}$ (note that the system always has a solution, which is the sample from which the measurements were obtained).

If we don't have any information on $\mathbf{x}$ we thus need at least $P=N$ measurements, and if we have them the problem is easy. As before instead, if there is a prior information on $\mathbf{x}$ we expect to need less measurements. Consider the case in which only $R = \rho N$ elements of $\mathbf{x}$ are non-zero, so that we have some sparsity measured by $\rho$. If we also knew \textit{which} elements of $\mathbf{x}$ are non-zero the problem would be equivalent to one with $R$ pixels, so the reconstruction is for sure impossible when $P < R$. On the other hand, having $P = R$ we can list all the $\binom{N}{R}$ vectors of dimension $N$ and sparsity $\rho$, try to solve all of these linear systems and in the end one of the choices will give us the solution (while the others will generically have no solution). Thus the sparsity information made the problem solvable for $P\geq R$.

So far, we have limited ourselves to considerations of whether or not the problem can be solved in certain regimes, but in practice this is certainly not enough: we want also to have an algorithm that solves the problem efficiently. More precisely we work in the $P,N,R \to \infty$ with $\alpha = P/N$ and $\rho = R/N$ finite and hope to find an algorithm that (when possible) solves the problem with polynomial complexity in $N$. Our reasoning from before shows the problem is impossible for $\alpha < \rho$, and if $\alpha \geq 1$ we are essentially inverting a matrix, so we have efficient algorithms for it. 
The region $\rho\leq\alpha<1$ is more interesting, as we would like to find the algorithm which allows us to solve the problem with the lowest $\alpha$. The enumerative procedure above solves the problem, but it is exponentially expensive in $N$, as

\begin{equation}
    \binom{N}{R} = \binom{N}{\rho N} \propto e^{N[-\rho \log{\rho} - (1-\rho)\log{(1-\rho)}]}.
\end{equation}

One can instead construct a non-optimal but polynomial algorithm by minimising an error function with $\textrm{L}_1$ regularisation, such as
\begin{equation}
    \min_{\mathbf{s}}\sum_{\mu=1}^P(y^\mu - \mathbf{F}^\mu \cdot \mathbf{s} )^2 + \lambda |\mathbf{s}|
\end{equation}
since the $\textrm{L}_1$ constraint reduces the effective allowed configurations for the problem. This is a convex function that can be minimized with efficient algorithms. Indeed, for fixed $\rho$ this procedure allows for perfect recovery for $\alpha>\alpha_{\textrm{L}_1}(\rho)$  
\cite{Donoho_2009, Kabashima_2009, Donoho_2009_universality} but it turns out that $\alpha_{\textrm{L}_1}(\rho)>\rho $, meaning that this algorithm obtained from a convex relaxation is sub-optimal. There exists an intermediate phase $\alpha_{\textrm{L}_1}(\rho)>\alpha>\rho$ where we know that we have enough information to reconstruct the sample fully, using exponential time enumeration, but the polynomial algorithm based on convex relaxation fails.

In the next sections we will give a procedure to describe the optimal error of the problem as a function of $\alpha, \,\rho$, as well as an algorithm that is derived as a consequence of the data model. But let us first put these preliminary examples in a broader context of inference, that encompasses some of the standard problems of machine learning.

\section{A general model}
In what follows we will study a general model of $N$ variables $\{x_i\}_{i\in N}$ with a factorised prior measure $P^\mathbf{x}(\mathbf{x}) = \prod_{i=1}^NP^\mathbf{x}_i(x_i)$, interacting through couplings $\Psi^\mu(\cdot)$ that can be seen as a set of $\mu =1,\ldots, P$ constraints, where $\Psi^\mu$ is a positive function of a linear combination of all the variables. The joint probability of that model for a given choice of $\mathbf{F}$ turns out to be 

\begin{equation}  \label{proba_general}
  P_{\mathbf{F}}(\mathbf{x}) = \frac{1}{\mathcal{Z}}\prod_{i=1}^NP^\mathbf{x}_i(x_i)\prod_{\mu=1}^P\Psi^\mu \left(\mathbf{F}^\mu \cdot\mathbf{x}\right).
\end{equation}
We assume $\Psi^\mu(\cdot) \geq 0$ while the measurement coefficients $F_{\mu i}$ are as usual quenched random variables drawn from a i.i.d. distribution with first two moments 
\begin{equation}
\mathbb{E}[F_{\mu i}] = 0\,, \qquad
\mathbb{E}[F_{\mu i}^2] = \frac{1}{N}.
\end{equation}
This ensures $F_{\mu i} \sim \mathcal{O}\left(\frac{1}{\sqrt{N}}\right)$ and, as a consequence, $\sum_i F_{\mu i}x_i \sim \mathcal{O}(1)$.  The analysis will be carried on in the thermodynamic limit $N,P\to\infty$, with $\alpha = P/N$ finite. 

Typically we will be interested in computing properties of this measure. For instance in computing the marginal probability of a single variable $m_i(x_i)=\sum_{\{x_j\},j\neq i}P(\bx)$ or the free entropy density $\Phi = \lim_{N\to\infty} \frac{1}{N} \mathbb{E}[\log{\mathcal{Z}}]$, where the average is taken over the choice of the $\mathbf{F}$ coefficients. The marginal probability will allow us to do a point estimation, meaning what we can find the best guess for $x_i$ in our problem. As an example, with $\textrm{L}_2$ error the best estimation is obtained by averaging over the posterior

\begin{equation}
    \hat x_i^{\textrm{opt}} = \langle x_i\rangle = \int x \,m_i(x)\, dx \ .
\end{equation}
The free entropy instead can give useful information on the complexity of the problem in the typical case since it concentrates in the $N\to \infty$ limit if $F_{\mu i}$ are generated with such an ensamble. It helps understand the phase diagram and it gives us exact asymptotics for the typical error \cite{guo2005mutual, Zdeborov__2016}, even if we will mostly focus on the marginals in the treatment.
\newline
Before proceeding, it is instructive to identify specific problems that we can examine as special instances of this overarching model. Notably, the initial two examples discussed in our introduction represent such special cases within this broader framework.

\subsection{Perceptron}
Let us consider again the problem of perceptron learning, as described in the introduction. Considering the description that we derived in (\ref{Perceptron_Bayesian}), we see that, whenever the regularization term is additive, $V(x)=\sum_i v(x_i)$, the perceptron problem can be cast into a special case of the general model, with
\begin{align}
  P_i^\bx(x_i)&=e^{-\beta V(x_i)}\\
  \Psi^\mu(\mathbf{F}^\mu \cdot\mathbf{x})&=e^{-\beta(y^\mu-\Phi(\mathbf{F}^\mu \cdot\mathbf{x}))^2}.
  \end{align}

The minimisation problem will be recovered in the zero-temperature limit $\beta\to\infty$ once sending $N,P\to\infty$. One can study the phase diagram for learning random classifications (for instance $y^\mu=\pm 1$ with probability $1/2$) \cite{gardner,gardner_derrida,krauth_mezard}, or for learning from an underlying "teacher's" rule where the $y^\mu$ are generated from a teacher perceptron with its own weights $x=t$, as mentioned in the introduction~\cite{Gyorgii}.

\subsection{Compressed sensing}
The model of compressed sensing that we described in the introduction is obviously an example of our general model, with
\begin{equation}
\Psi^\mu(\mathbf{F}^\mu \cdot\mathbf{x}) = e^{-\frac{1}{2\Delta^2}(y^\mu - \mathbf{F}^\mu \cdot\mathbf{x})^2} 
\end{equation}
The distribution $P_i^\mathbf{x} $ should impose a level of sparsity. For instance one can use a Gauss-Bernoulli prior of the form 
\begin{equation}
    P_i^\mathbf{x}(x_i) = (1-\rho)\delta(x_i) + \rho \frac{e^{-\frac{x_i^2}{2}}}{\sqrt{2\pi}}.
\end{equation}

\subsection{Generalized linear regression}
This is a simple generalization of compressed sensing. Imagine you have $P$ patients and patient $\mu$ has a level of expression of a disease $y^\mu$. On the other hand, for each of them one has measured the values of $N$ parameters, like the age, the concentration of some molecules etc.; the set of measurements being summarized in a $P\times N$ matrix $F$. One might want to do some regression and find the best linear combination of the parameters that fits the data $y$. An exact fit would be found if there exists a set of values $x_i$, $i\in \{1,...,N\}$ such that $\forall \mu:\ \ y^\mu =\mathbf{F}^\mu \cdot\mathbf{x}$. In general, one can expect that the measured $y^\mu$ is a noisy version of $\mathbf{F}^\mu \cdot\mathbf{x} $. This can be expressed by assuming the existence of a noisy channel 
$ P_c(y^\mu|\mathbf{F}^\mu \cdot\mathbf{x})$. Then the generalized linear regression amounts to finding the most probable value of the $x_i$ where the probability law is
\begin{equation}
   P(x)=\frac{1}{Z}\;\prod_i P_i^\mathbf{x} \;\prod_\mu P_c(y^\mu|\mathbf{F}^\mu \cdot\mathbf{x})
\end{equation}
where $P_i^\mathbf{x} $ includes our prior knowledge on $x_i$.

   \subsection{The Hopfield model}
   Hopfield's model of associative memory \cite{Hopfield} is based on a set of binary spins (corresponding to neuron activities) interacting by pairs, with an energy function $E[\mathbf{x}] = -\frac{1}{2}\sum_{i,j}J_{ij}x_ix_j$. It depends on a symmetric matrix of coupling constraints $J_{ij}$ with zero diagonal elements ($J_{ii}=0$), aaimplying there are no self-interactions). In that setting, one can store as stable fixed points a number $P$ of random patterns $\xi^\mu$, which are binary spin configurations $\bxi^\mu=\{\xi_1^\mu,...,\xi_N^\mu\}$, by choosing the $J_{ij}$ as 
\begin{equation}
J_{ij}=\frac{1}{N}\sum_{\mu=1}^P \xi_i^\mu \xi_j^\mu
\end{equation}
that is called the \textit{Hebb rule}. In practice, once defined the time evolution rule of the system
\begin{equation}
    x_i(t+\Delta t) = \text{sign}\left(\sum_j J_{ij} \,x_j(t)\right)
\end{equation}
one can verify the stability of each stored patterns as fixed points of the dynamics if the state of the system is in perfect coincidence with the pattern $\mu$ at time $t$, i.e. $x_i(t) = \xi^\mu_i \,\forall i$.
For what is of interest to us in the discussion, however, the probability distribution of the spins at inverse temperature $\beta$ is
\begin{equation}
    P_J(\mathbf{x}) = \frac{1}{\mathcal{Z}}\prod_i \rho(x_i) e^{(\beta/2)\sum_{i,j=1}^N J^\mu_{ij} x_i x_j}
    =\frac{1}{\mathcal{Z}}\prod_i\rho(x_i) e^{\frac{\beta}{2N}\sum_{\mu=1}^N (\bxi^\mu . \bx)^2}
\end{equation}
where $\rho(x)$ is the Ising measure, enforcing $x=\pm 1$ with probability $1/2$.
This is again of our general type with a function $\Psi(t)=e^{(\beta/2)t^2}$. Recent generalizations of the Hopfield model, with a larger storage capacity, use rather $\Psi(t)=e^{(\beta/N^{k-1}) \; t^k}$  with $k>2$ \cite{Gardner_k}, or even exponential functions~\cite{Krotov_Hopfield,Ramsauer,Lucibello_MM}.

Note that in the large $P$ limit the  elements of the coupling matrix  become independent and one recovers the well-known Sherrington-Kirkpatrick model \cite{SK}, which is thus a limiting case of our general model.

\subsection{Code Division Multiple Access (CDMA)}

In wireless communication, one uses a communication system from the user devices to a base station where the information is coded in order to share the same communication medium. 
In CDMA, this sharing is represented as
\begin{equation}
  \by=F \bx^0+{\boldsymbol{\eta}}\ ,
\end{equation}
where $\bx^0\in\mathbb{R}^N$ are the information symbols
to be sent by $N$ devices, $F\in\mathbb{R}^{P\times N}$
is the mixing matrix, and $\by\in\mathbb{R}^P$ is the signal received at the base station. Each device $i\in\{1,...,N\}$ uses its own mixing vector ${\bf{F_i}}=\{F_{\mu i}\}$ that characterizes what it sends through each channel $\mu$. Here we have supposed that the communication between the user and the base has an additive noise $\boldsymbol{\eta}$, but more general models can be considered.

Having received $\by$, the base must infer from  the received messages $\by$ and the knowledge of the mixing matrix $F$, what where the information symbols $x_i$ sent by each of the users. 
If these information symbols are generated i.i.d. from a probability distribution
$p(\bx)=$, and the $P$ components of the noise are i.i.d., distributed with a distribution $\rho$, the Bayesian inference of the $x_i$ amounts to studying the posterior probability
\begin{equation}
  p(\bx\mid\by,F)=\frac{1}{Z(\by,F)} \; \prod_{i=1}^Np(x_i)\; \prod_{\mu=1}^P\rho(y^\mu-\sum_i F_{\mu i} x_i)
\end{equation}
which is again an instance of our general model. The statistical physics formulation and solution of CDMA is due to Tanaka \cite{Tanaka2001,Tanaka2002} and Kabashima \cite{Kabashima}. A nice review can be found in \cite{KT-FarBeyond2023}.

\section{Belief propagation: general introduction}
\begin{figure}
\centering
\begin{minipage}{.5\textwidth}
  \centering
  \includegraphics[width=.7\linewidth]{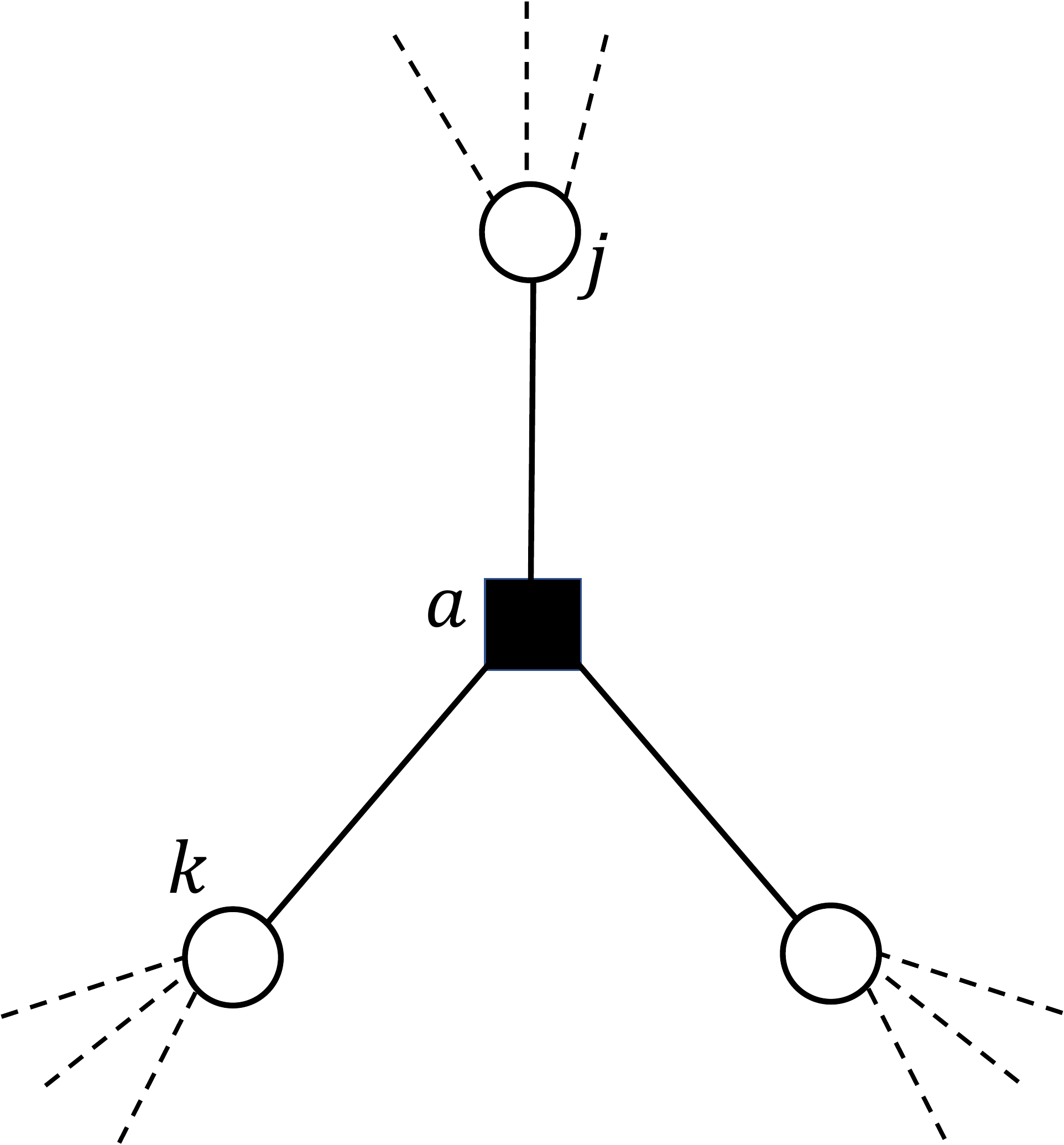}
\end{minipage}%
\begin{minipage}{.5\textwidth}
  \centering
  \includegraphics[width=.7\linewidth]{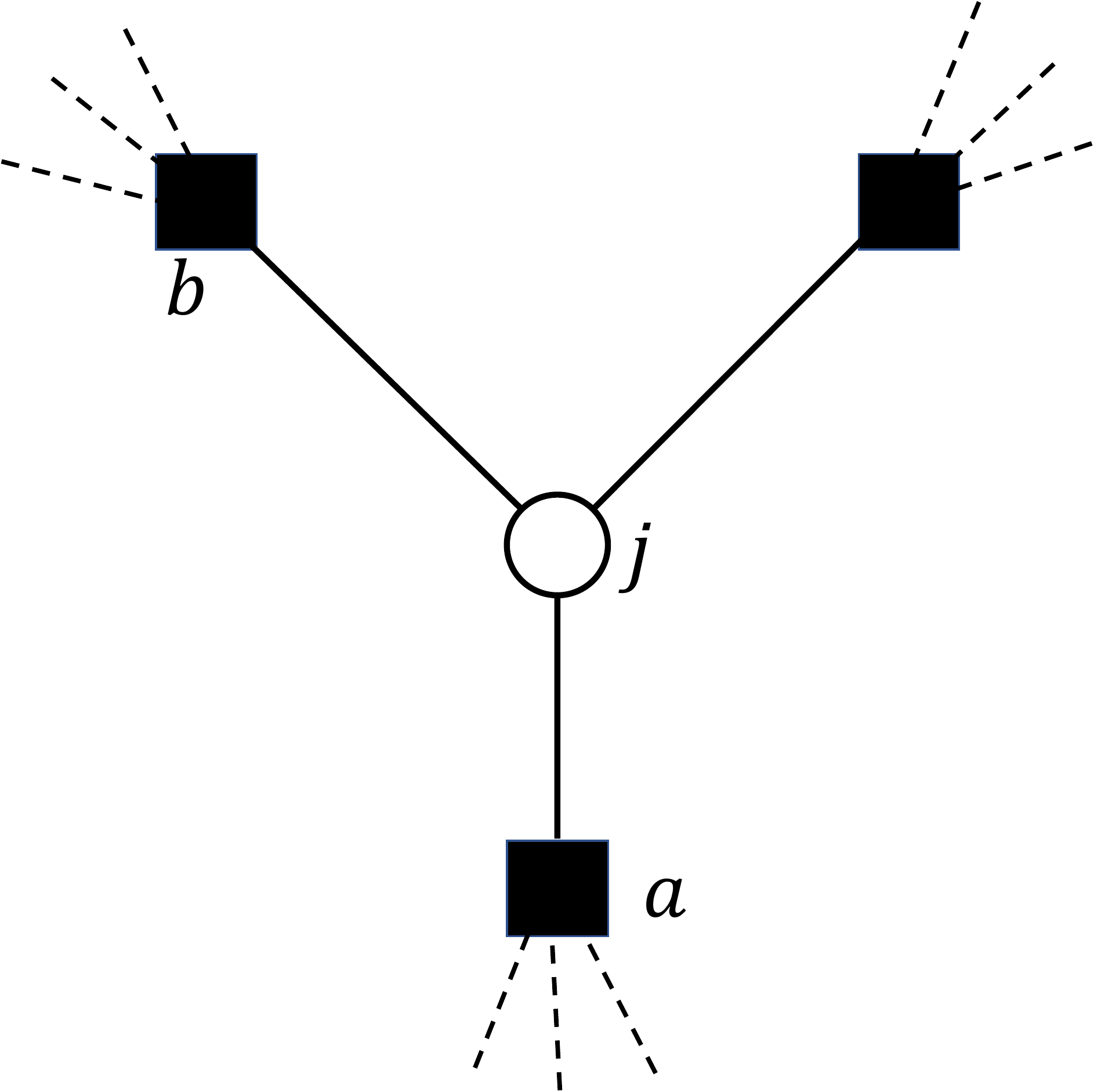}
\end{minipage}
\caption{
Pictorial representation of the messages.
}\label{fig:BP}
\end{figure}

In this section we give a short introduction to the study of general graphical models using 
mean field equations called belief propagation(BP) (for a more extensive presentation, see \cite{MezMon}).
Consider a set of random variables $x_i \in \chi$ where $\chi$ is a finite space. We define a factor graph, that is a bipartite graph whose nodes are either \textit{variable nodes} associated to variables $x_i$ or \textit{factor nodes} associated to the probability factors $\psi_a$. If a factor acts on a variable the corresponding nodes are connected as in figure \ref{fig:BP}. In formulas we are asking that $\mathbf{x}$ obeys the joint probability distribution

\begin{equation} \label{factorg}
  P(\mathbf{x}) = \frac{1}{Z} \prod_{a=1}^M \psi_a(\mathbf{x}_{\partial a})
\end{equation}
being $\mathbf{x}_{\partial a} = \{x_i\,|\,i\in\partial a\}$ the set of variables appearing in the factor $\psi_a$ which is a positive function of the the variables $\{x_i\},\ i\in \partial a$ to which it is connected. Of course the graph does not represent the full value of the probability distribution \eqref{factorg} but it says what variables appear in what factors and complemented with the information about the kind of function implemented in each factor completely characterize the problem. 

For the derivation of the mean field equations we suppose the factor graph to be a tree, so our computation will be exact. This means the graph is acyclic, i.e. there is no way to start at a node, move along the edges, and return to the same node without retracing your steps. There are several examples where this occurs: in coding theory, for instance, linear block codes like Hamming codes can be represented as a tree-structured factor graph. In that case the variables are the code symbols, and the factors represent the parity-check relations between the symbols. In any case, these equations can be used in more general settings, and their validity can be controlled when the factor graph is locally-tree like (modulo possible effects of ``replica symmetry breaking''). 

Now, suppose we are interested in computing the marginal $m_j(x)$

\begin{equation}
        m_j(x) = \sum_{\{x_i\}_{i\neq j}}P(\mathbf{x}).
\end{equation}

A priori, to compute this quantity, supposing we are dealing with a binary system $\{x_i\}_{i=1}^N =\pm 1$, we need to implement a number $2^{N-1}$ of sums. The claim is that there is a smarter way to do that which instead of an exponential operation is just linear in the size of the system. To explain how, let us define messages, or beliefs, on the factor graph. They are probability distributions with support on $\chi$ associated to the edges of the graph. In the following we will write $j\to a$ to indicate that we cut the edge $(j, a)$ from the graph. Define the messages $m_{j\to a}$, $\hat m_{a\to j}$ from the variables to the factor and from the factor to the nodes. They are the marginal  probability distributions of the variable $x_j$ in the "cut" factor graph from the variable and the factor side respectively.
In tree factor graphs, it is possible to write self-consistent relations between the messages, commonly called Belief Propagation (BP) equations (figure \ref{fig:BP})
\begin{equation}
\begin{split}
    m_{j\to a}(x_j) &\propto \prod_{b\in \partial j\setminus a} \hat m_{b\to j}(x_j)\\
    \hat m_{a\to j}(x_j) &\propto \sum_{\{\mathbf{x}_{\partial a\setminus j}\}} \psi_a(\mathbf{x}_{\partial a}) \prod_{k\in \partial a\setminus j}  m_{k\to a}(x_k) \\
\end{split}
\end{equation}
where the symbol $\propto$ means ``proportional to'' (all messages are probabilities and should thus be normalised). 
If $\partial j \setminus a$ is the empty we have

\begin{equation}
    m_{j\to a}(x_j) = \frac{1}{|\chi |} \,,
\end{equation}
similarly, for an empty $\partial a \setminus j$:
\begin{equation}
    \hat m_{a\to j}(x_j) \propto \psi_a(x_j) \,.
\end{equation}The marginal probability $m_j$ has a simple expression in terms of the messages

\begin{equation}
    m_j(x) \propto \prod_{b\in \partial j} \hat m_{b\to j}(x_j).
\end{equation}

It is also possible to derive the following expression for the free entropy \cite{MezMon}

\begin{equation}
  \Phi = \log{\mathcal{Z}} = \sum_a \log{\mathcal{Z}_a} + \sum_i \log{\mathcal{Z}_i} -\sum_{(i,a)\in E} \log{\mathcal{Z}_{ia}}  
\end{equation}
where
\begin{equation}
\begin{split}
  \log{\mathcal{Z}_a} &= \sum_{\{\mathbf{x}_{\partial a}\}} \prod_{j\in \partial a} m_{j\to a}(x_j)\, \psi_a(\mathbf{x}_{\partial a})\\
  \log{\mathcal{Z}_i} &= \sum_{\{x_i\}} \prod_{a\in \partial i} \hat m_{a\to i}(x_i)\\
  \log{\mathcal{Z}_{ia}} &= \sum_{\{x_i\}} m_{i\to a}(x_i)\, \hat m_{a\to i}(x_i).\\
\end{split}
\end{equation}
Clearly $\Phi_i = \ln {\mathcal{Z}_i}$ is a site
term, that measures the free energy change when the site $i$ and all its edges are added; $\Phi_a = \log{\mathcal{Z}_a}$ instead is a local interaction term that gives the free energy change when the function node $a$ is added to the factor graph. Finally $\Phi_{ia} =  \log{\mathcal{Z}_{ia}}$ is an edge term, which takes into account the fact that in adding vertex $i$ and $a$, the edge $(i, a)$ is counted twice. The BP equations can be used as a fixed point scheme: we start by randomly sampling the initial messages $\hat{m}_{a\to i}^0$ and then we iterate
\begin{equation}
\begin{split}
    m_{j\to a}^{t+1}(x_j) &= \prod_{b\in \partial j\setminus a} \hat m_{b\to j}^t(x_j)\\
    \hat m_{a\to j}^{t+1}(x_j) &= \sum_{\{\mathbf{x}_{\partial a\setminus j}\}} \psi_a(\mathbf{x}_{\partial a}) \prod_{k\in \partial a\setminus j}  m_{k\to a}^{t+1}(x_k). \\
\end{split}
\end{equation}
If the factor graph is a tree the number of iterations to converge from the leaves towards the centre is exactly the diameter of the tree. 
\section{Belief propagation for the General Model: "Reduced-BP"}
\begin{figure}
    \centering
    \includegraphics[width=6cm]{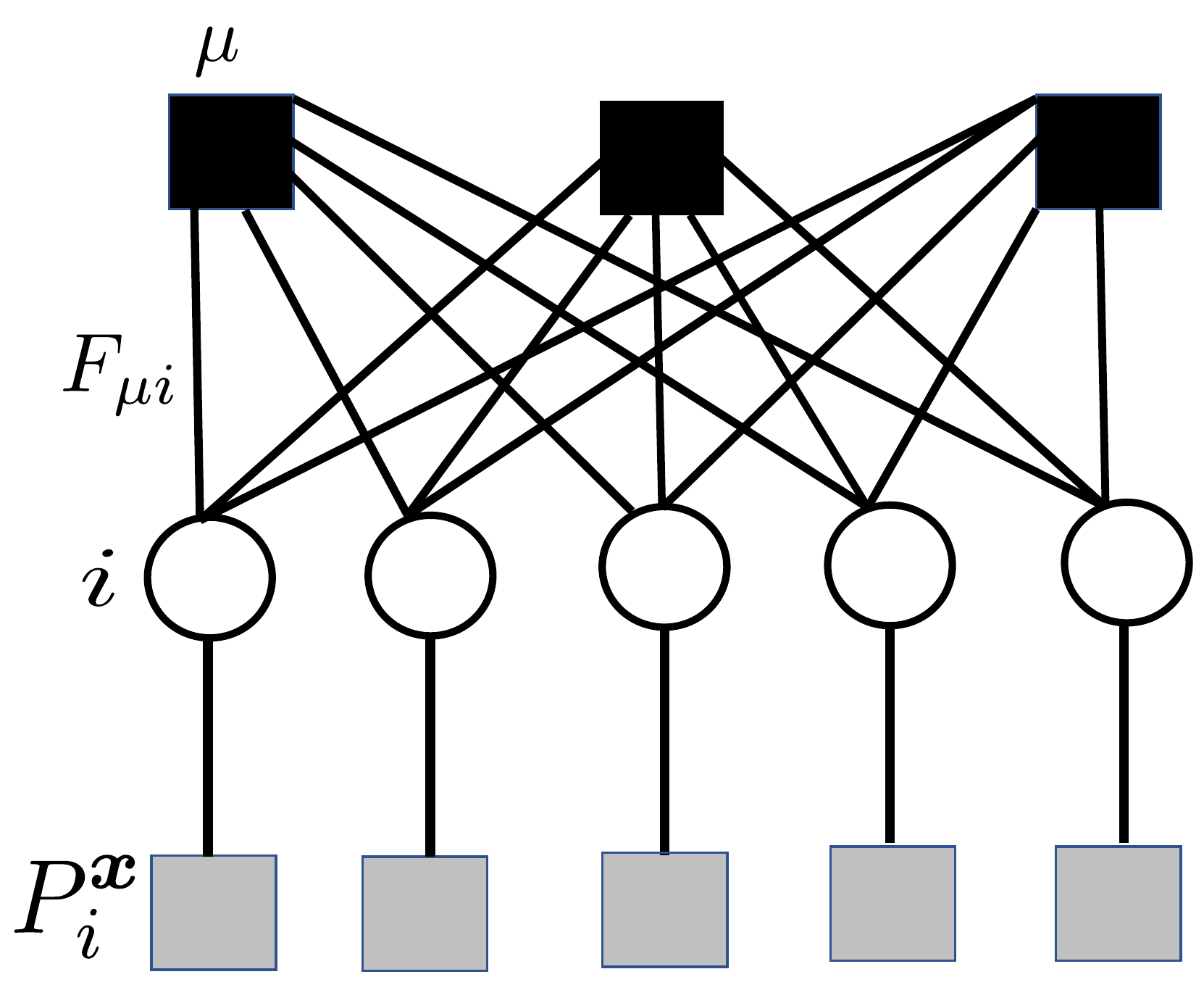}
    \caption{The general model of a fully-connected graph.}
    \label{fig:general}
\end{figure}
Let us write the BP equations for the general model (\ref{proba_general}), that is a fully connected graph in which each interaction involve all the variable nodes as in figure \ref{fig:general}.
There are two kinds of factors: $P_i^{\mathbf{x}}(x_i)$, that depends only on a single variable, and $\Psi^\mu(\mathbf{F}^\mu \cdot \mathbf{x})$, that couples different variables. We rewrite the BP equations as

\begin{align} \label{hatmueq}
    m_{i\to \mu}(x_i) &= P_i^x(x_i)\prod_{\nu (\neq\mu)} \hat m_{\nu\to i}(x_i)\\
    \hat m_{\mu \to i}(x_i) &= \int \prod_{j (\neq i)}[dx_j m_{j\to \mu}(x_j)] \Psi^\mu \left(F_{\mu i} x_i + \sum_{j(\neq i)} F_{\mu j} x_j\right).
\end{align}
There are two main differences between these equations and the ones in the previous section: one  is the use of continuous variables instead of discrete ones, the other one is the fact that the interacting factors $\Psi^\mu$ typically involve a linear combination of many variables.
It would seem that these equations are not tractable anymore, as they involve high-dimensional integrals which are hard to compute in practice. However, it turns out that, in some cases, the messages can be simplified in the large $N$ limit, using the central limit theorem. For this we basically need that the number of variables appearing in each factor  $\Psi^\mu$ diverges in the large $N$ limit, and that the coefficients $F_i^\mu$  be balanced, i.e. all of the same order and with balanced signs. We shall reason here for simplicity in the case where  $F_i^\mu$ are i.i.d. random values with mean zero and variance $1/N$.

In this case, equation (\ref{hatmueq}) contains the sum $u_{\mu\to i}=\sum_{j\neq i} F_{\mu j} x_j$, in which  the variables $x_j$ are independent, each one being distributed according to $m_{j\to \mu}$.  When $N \to \infty$ becomes a Gaussian random variable.

In order to characterize it, let us introduce the first and second moments of messages $m_{i\to\mu}$
\begin{equation} \label{a_r_BP}
\begin{split}
    a_{i\to \mu} &= \int dx\, m_{i\to \mu}(x)\,x\\
    v_{i\to \mu} &= \int dx\, x^2 m_{i\to \mu}(x) - a_{i\to \mu}^2.\\
\end{split}
\end{equation}
Then the mean and variance of $u_{\mu\to  i}$ , denoted respectively 
$\omega_{\mu\to i}$ and $V_{\mu\to i}$, are  given by 
\begin{equation}
\begin{split}  \label{omega-V-def}
    \omega_{\mu\to i} &=\mathbb{E}\left[u_{\mu\to i}\right]= \sum_{j\neq i} F_{\mu j} a_{j\to \mu}\\
    V_{\mu\to i} &=\mathbb{E} \left[u_{\mu\to i}^2\right]- \mathbb{E}\left[ u_{\mu\to i}\right]^{\,2} =  \sum_{j\neq i} F_{\mu j}^2 v_{j\to \mu}\,.\\
  \end{split}
\end{equation}
The messages $\hat m_{\mu \to i}(x_i) $ then simplify to
\begin{equation} \label{hatmueq_2}
\hat m_{\mu \to i}(x_i)\propto \int \frac{du_{\mu\to i}}{\sqrt{2\pi V_{\mu\to i}  }} e^{-(u_{\mu\to i}-\omega_{\mu\to i})^2/2 V_{\mu\to i}} \,\Psi^\mu (F_{\mu i} x_i + u_{\mu\to i}) 
\end{equation}
As $F_{\mu i}=O(1/\sqrt{N})$, this equation can be expanded to second order in $ F_{\mu i} x_i $ and re-exponentiated. An easy way to perform this step is to introduce the Fourier transform $\tilde \Psi^\mu(k)$ of the factor $\Psi^\mu$

\begin{equation}
    \tilde \Psi^\mu(k)= \int dx\, \Psi^\mu(x) e^{-ikx}
\end{equation}
Then one gets
\begin{align} \label{mess-step1}
  \hat m_{\mu \to i}(x_i) &\propto \int \frac{dk}{2\pi}\int \frac{du_{\mu\to i}}{\sqrt{2\pi V_{\mu\to i}  }} e^{-[(u_{\mu\to i}-\omega_{\mu\to i})^2]/[2 V_{\mu\to i} ]} \tilde \Psi^\mu (k) e^{ik (F_{\mu i} x_i + u_{\mu\to i})}
    \nonumber\\
  &\propto \int \frac{dk}{2\pi} e^{- V_{\mu\to i}k^2/2+ ik (F_{\mu i} x_i +\omega_{\mu\to i})} \tilde \Psi^\mu (k).
\end{align}
We can now expand this expression to second order in $F_{\mu i}$ (which is of order $1/\sqrt{N})$, and re-exponentiate it. This shows that the message takes a Gaussian form in the large $N$ limit, a consequence of the central limit theorem. Precisely, one gets
\begin{equation}
  \hat m_{\mu \to i}(x_i)\propto \exp\left[F_{\mu i} x_i G_1^\mu(V_{\mu\to i},\omega_{\mu\to i})+\frac{1}{2}
  F_{\mu i}^2 x_i^2 G_2^\mu(V_{\mu\to i},\omega_{\mu\to i})  \right]
\end{equation}
where we have introduced the functions 
\begin{equation}
  \begin{split}
  G_0^\mu(V,\omega)&= \int \frac{dk}{2\pi} e^{- V k^2/2+ ik \omega  } \tilde \Psi^\mu (k) \\
  G_1^\mu(V,\omega)&=\frac{1}{G_0^\mu (V,\omega)}\frac{d G_0^\mu (V,\omega)} {d\omega} \\
  G_2^\mu(V,\omega)&= \frac{d G_1^\mu (V,\omega)} {d\omega} =\frac{1}{G_0^\mu (V,\omega)}\frac{d^2 G_0^\mu(V,\omega)}{d\omega^2}-[G_1^\mu(V,\omega)]^2
  \end{split}
\end{equation}
we finally arrive at
\begin{equation} \label{rBP1}
    \hat m_{\mu \to i}(x_i) = \frac{1}{\hat{\mathcal{Z}}_{\mu\to i}} e^{B_{\mu \to i} x_i - \frac{1}{2}A_{\mu \to i}x_i^2}
  \end{equation}  
with
\begin{equation}
\begin{split}
  A_{\mu \to i} &= -F_{\mu i}^2 G_2^\mu(V_{\mu \to i},\omega_{\mu \to i})\\
  B_{\mu \to i} &=  F_{\mu i} G_1^\mu(V_{\mu \to i},\omega_{\mu \to i})\\
  \end{split}
  \end{equation}  
Inserting this expression into the equation for the message $m_{i\to\mu}$ we find
\begin{equation}
    m_{i\to \mu}(x_i) = \frac{1}{\mathcal{Z}_{i\to \mu}} P_i^x(x_i) e^{x_i \sum_{\nu\neq\mu}B_{\nu\to i} - (x_i^2/2) \sum_{\nu\neq\mu}A_{\nu\to i}}
  \end{equation}
For convenience we rewrite $m_{i\to\mu}$ as
\begin{equation}  \label{rBP2}
    m_{i\to \mu}(x_i) = \frac{1}{\mathcal{Z}_{i\to \mu}} P_i^x(x_i) e^{
    -\frac{(R_{i\to\mu}-x_i)^2}{2\Sigma_{i\to \mu}}}
\end{equation}
where we defined $\Sigma_{i\to\mu}$, $R_{i\to\mu}$:

\begin{equation}
    \Sigma_{i\to\mu} = \left[\sum_{\nu\neq\mu}A_{\nu\to i}\right]^{-1}\,, \qquad R_{i\to\mu} = \left[\sum_{\nu\neq\mu}A_{\nu\to i}\right]^{-1}\left[\sum_{\nu\neq\mu}B_{\nu\to i}\right]
\end{equation}

  The equations (\ref{rBP1}, \ref{rBP2}), where $\mathcal{Z}_{i\to \mu}$, $\hat{\mathcal{Z}}_{\mu\to i}$ are normalisation constants, provide a set of self-consistent equations for the messages $A_{\mu \to i}$, $B_{\mu \to i}$ and $\omega_{\mu\to i}$, $V_{\mu\to i}$. We can try to solve them by iteration; we thus have a fixed point scheme in term of the \textit{moments} of the messages.

  The equations were first written in the case of the perceptron in \cite{Mezard89}. They are actually the cavity equations obtained by applying the cavity method (initially introduced to deal with the SK model \cite{MPV_cavity}, to the perceptron, following the strategy developed for Hopfield's model \cite{mezard_parisi_virasoro_2004}). As an algorithm, these equations have been introduced for CDMA in \cite{Kabashima}, and developed under the name of  reduced BP (r-BP) in many works \cite{Rangan, Guo_2006}. While r-BP is surely much more effective than applying the BP equations blindly, for many applications it is still an "expensive" algorithm, as it uses $\mathcal{O}(N^2)$ messages as the number of edges in the graph. In the next part we will show how it can be simplified into an algorithm on the nodes, which will have complexity $\mathcal{O}(N)$. This is the step that one takes when going from the cavity equations (which use 'cavity messages', i.e. magnetizations defined in absence of a neighboring site) to TAP equations (which use the standard magnetizations).

\section{"TAPification": From reduced-BP to Approximate Message Passing}

Let's be tempted for a moment by the following, very intriguing, approximation $\omega_{\mu\to i}\approx\omega_\mu$, $V_{\mu\to i}\approx V_\mu$

\begin{equation}
\begin{split}
        \omega_{\mu\to i} = \sum_{j\neq i} F_{\mu j} a_{j\to \mu} &\approx \sum_j F_{\mu j} a_{j\to \mu} = \omega_\mu\\
        V_{\mu\to i} = \sum_{j\neq i} F_{\mu j}^2 v_{j\to \mu} &\approx \sum_j F_{\mu j}^2 v_{j\to \mu} = V_\mu.\\
\end{split}
\end{equation}
This would be enough to obtain a $\mathcal{O}(N)$ algorithm, as we can redefine $\Sigma_{i\to\mu}$, $R_{i\to\mu}$ as $\Sigma_i$, $R_i$, hence the messages $m_{i\to\mu}$ will just depend on the nodes and be replaced by $m_i$, which are then also the marginals

\begin{equation}
    m_i(x) = \frac{1}{\mathcal{Z}_i} P_i^\mathbf{x}(x) e^{-\frac{(x-R_i)^2}{2\Sigma_i}}.
\end{equation}
It is in fact possible to do this simplification, but we need to be careful. By keeping correctly the leading terms in $1/N$ we will obtain a correction, called Onsager reaction term. Let's look at how this reaction term appears. We first consider $\Sigma_{i\to\mu}$

\begin{equation}
    \Sigma_{i\to\mu} = \left[-\sum_{\nu\neq\mu}F_{\nu i}^2 G_2^\nu(V_{\nu \to i},\omega_{\nu \to i})\right]^{-1}\approx\left[-\sum_\nu F_{\nu i}^2 G_2^\nu(V_\nu,\omega_\nu)\right]^{-1}.
\end{equation}
We thus define $\Sigma_i$
\begin{equation}  \label{Sigmai-def}
  \Sigma_i = \left[-\sum_\mu F_{\mu i}^2 G_2^\mu(V_\mu,\omega_\mu)\right]^{-1}.
\end{equation}
Notice that at order $1/N$ we have
\begin{equation}
    B_{\mu \to i} = F_{\mu i} G_1^\mu(V_{\mu \to i},\omega_{\mu \to i}) \approx F_{\mu i} G_1^\mu(V_\mu,\omega_\mu) - F_{\mu i}^2 a_i \frac{d G_1^\mu (V_\mu,\omega_\mu)} {d\omega}.
\end{equation}
We can now expand $R_{i\to\mu}$ to leading order

\begin{equation}
\begin{split}
    R_{i\to\mu} = \left[-\sum_{\nu\neq\mu}F_{\nu i}^2 G_2^\nu(V_{\nu \to i},\omega_{\nu \to i})\right]^{-1}&\left[\sum_{\nu\neq\mu}F_{\mu i} G_1^\mu(V_{\mu \to i},\omega_{\mu \to i}) - F_{\mu i}^2 a_i \frac{d G_1^\mu (V_\mu,\omega_\mu)} {d\omega}\right]\approx \\
    &\qquad\qquad\qquad\qquad\approx a_i + \Sigma_i\sum_{\nu}F_{\nu i} G_1^\nu(V_{\nu},\omega_{\nu})\\
\end{split}
\end{equation}
so we have that
\begin{equation}  \label{Ri-def}
  R_i = a_i + \Sigma_i\sum_{\mu}F_{\mu i} G_1^\mu(V_{\mu},\omega_{\mu}).
\end{equation}
It is convenient to introduce the first two moments of $m_i$ at leading order:
\begin{equation} \label{ai-vi-def}
\begin{split}
    a_i &= \int dx\, m_i(x)\,x \approx \int x\,\mathcal{D}m_i(R_i,\Sigma_i)\\
    v_i &= \int dx\, x^2 m_i(x) - a_i^2 \approx \int x^2\,\mathcal{D}m_i(R_i,\Sigma_i)- a_i^2 \approx \Sigma_i\frac{\partial a_i}{\partial R_i}\\
  \end{split}
\end{equation}
where $\mathcal{D}m_i(R_i,\Sigma_i)$ is the message measure:
\begin{equation} \label{eq:ref_measure}
    \mathcal{D}m_i(R_i,\Sigma_i) = \,\frac{1}{\mathcal{Z}_i(R_i, \Sigma_i)}dx\,P^\mathbf{x}_i(x) e^{-\frac{(x-R_i)^2}{2\Sigma_i}}.
\end{equation}
We can finally write the leading order of $a_{i\to\mu}$, $v_{i\to\mu}$
\begin{equation}
\begin{split}
    a_{i\to\mu} &\approx a_i - F_{\mu i}  G_1^\mu(V_{\mu},\omega_{\mu}) \Sigma_i\frac{\partial a_i}{\partial R_i} = a_i - F_{\mu i}  G_1^\mu(V_{\mu},\omega_{\mu}) v_i\\
    v_{i\to\mu} &\approx v_i.
\end{split}
\end{equation}
As a consequence we obtain the relations
\begin{equation}  \label{omegamu-Vmu-def}
\begin{split}
    \omega_\mu &= \sum_j  F_{\mu j} a_j - G_1^\mu(V_{\mu},\omega_{\mu})\sum_jF_{\mu j}^2 v_j\\
    V_\mu &= \sum_j F_{\mu j}^2 v_j.
  \end{split}
\end{equation}
This concludes the derivation of the TAP equations, which are a set of auto-coherent equations relating the variables $a_i,v_i,\Sigma_i, R_i$ and $\omega_\mu,V_\mu$ given in (\ref{ai-vi-def},\ref{Sigmai-def},\ref{omegamu-Vmu-def}). These are site variables, and therefore the total number of variables (and of equations) scales linearly with $N$.

One can try to solve these equations above by iteration, which can be done in the following way

\begin{equation}
\begin{split}
    V_\mu^{t+1} &= \sum_j F_{\mu j}^2 v_j^t\\
    \omega_\mu^{t+1} &= \sum_j  F_{\mu j} a_j^t - G_1^\mu(V_{\mu}^t,\omega_{\mu}^t)\sum_jF_{\mu j}^2 v_j^t\\
    \Sigma_i^{t+1} &= \left[-\sum_\mu F_{\mu i}^2 G_2^\mu(V_\mu^{t+1},\omega_\mu^{t+1})\right]^{-1}\\
    R_i^{t+1}&= a_i^t + \Sigma_i\sum_{\mu}F_{\mu i} G_1^\mu(V_{\mu}^{t+1},\omega_{\mu}^{t+1})\\
    a_i^{t+1} &= \int x\,\mathcal{D}m_i(R_i^{t+1},\Sigma_i^{t+1})\\
    v_i^{t+1} &= \int x^2\,\mathcal{D}m_i(R_i^{t+1},\Sigma_i^{t+1})- (a_i^{t+1})^2.\\
\end{split}
\end{equation}

This is a  $\mathcal{O}(N)$ algorithm for solving compressed sensing, usually called Approximate Message Passing (AMP) \cite{Donoho_2009, Rangan}. The specific choice of time indexes is crucial for convergence \cite{Bolthausen}.

\section{State evolution and phase transitions}
We shall present this section in the specific case of compressed sensing, 
as we believe it paints a clearer picture.
In the compressed sensing case the r-BP equations read
  
\begin{equation}
    \hat m_{\mu \to i}(x_i) = \frac{1}{\hat{\mathcal{Z}}_{\mu\to i}} e^{B_{\mu \to i} x_i - \frac{1}{2}A_{\mu \to i}x_i^2}
\end{equation}
with:
\begin{equation}
\begin{split}
    A_{\mu \to i} &= \frac{F_{\mu i}^2}{\Delta^2 + V_{\mu\to i}}\\
    B_{\mu \to i} &= \frac{F_{\mu i}(y_\mu - \omega_{\mu\to i})}{\Delta^2 + V_{\mu\to i}}\\
\end{split}
\end{equation}
$\omega_{\mu\to i}$ and $V_{\mu\to i}$ are the first two moments of $\sum_{j\neq i} F_{\mu j}x_j$ \textit{averaged over the messages}, as defined in (\ref{omega-V-def}).
We would like to analyse the asymptotic performance of AMP in the $N\to\infty$ limit. In particular, we would like to study the asymptotic error $E$

\begin{equation}
    E = \lim_{N\to\infty}\frac{1}{N}\sum_{i=1}^N (\hat x_i - s_i)^2\,,
\end{equation}
where $\hat x$ is the estimator obtained by running the algorithm. 
As a matter of fact the tracking of this metric is already embedded in the equations, but we need to exploit the way in which $y_\mu$ is generated. First, we define $z_{\mu\to i}=\sum_{j\neq i} F_{\mu j} s_j$ in analogy with $u_{\mu\to i}$. Then we have:

\begin{equation}
    y_\mu = z_{\mu\to i} + F_{\mu i} s_i + \eta_\mu
\end{equation}

In fact we can take the r-BP equations and notice that
\begin{equation} \label{E_concentration}
\begin{split}
    \mathbb{E}\left[\left(z_{\mu\to i} - \omega_{\mu\to i}\right)^2\right] =& \sum_{\substack{j\neq i \\ k\neq i}} \mathbb{E}[F_{\mu j}F_{\mu k}] \left(a_{j\to \mu}a_{k\to \mu} - 2s_ja_{k\to \mu} + s_js_k\right)=\\
    =&\frac{1}{N} \sum_{j\neq i} \left(a_{j\to i}^2 -2s_ja_{j\to i} + s_j^2\right)=\\
    =&\frac{1}{N} \sum_{i=1}^N \left(\hat{x}_i^2 -2s_i\hat{x}_i + s_i^2\right)=\\
    =&\frac{1}{N}\sum_{i=1}^N (\hat x_i - s_i)^2
\end{split}
\end{equation}
where we used that $a_{i\to \mu}$ and $v_{i\to \mu}$ are, up to small corrections, the mean and variance of the estimates $\hat x_i$, as well as the weak coupling hypothesis on $F_{\mu i}$. In the $N\to\infty$ limit the quantity above will converge to the error $E$. In particular, $\left(z_{\mu\to i} - \omega_{\mu\to i}\right)^2$ will concentrate on $E$ in the limit of large N, as the variance is subleading. It will be convenient in the following to introduce another parameter of interest, namely the squared distance of two components of $\hat{x}$
\begin{equation}
    v = \frac{1}{2N^2} \sum_{i,j=1}^N (\hat x_i - \hat x_j)^2\,.
\end{equation}
In particular $V_{\mu\to i}$ will concentrate on $v$
\begin{equation}
    \mathbb{E}[V_{\mu\to i}] = \sum_{j\neq i} \mathbb{E}[F_{\mu j}^2] v_{j\to \mu} = \frac{1}{N}\sum_{j\neq i} v_{j\to \mu} \approx v\,.
\end{equation}
We are ready for the main calculation. Let's look at the asymptotics of $\sum_\mu A_{\mu\to i}$, $\sum_\mu B_{\mu\to i}$:
\begin{equation}
    \sum_\mu A_{\mu\to i} = \sum_\mu\frac{F_{\mu i}^2}{\Delta^2 + V_\mu} = \frac{\alpha}{\Delta^2 + v}
\end{equation}

\begin{equation}
    \sum_\mu B_{\mu \to i} = \sum_\mu\frac{F_{\mu i}(y_\mu - \omega_{\mu\to i})}{\Delta^2 + V_{\mu\to i}} \approx \sum_\mu\frac{F_{\mu i}(y_\mu - \omega_{\mu\to i})}{\Delta^2 + v}\,.
\end{equation}
The expression we just derived has a hidden dependence on $F_{\mu i}$ inside $y_\mu$, so to proceed we insert its definition:
\begin{equation}
\begin{split}
    \sum_\mu F_{\mu i}(y_\mu - \omega_{\mu\to i}) &= \sum_\mu F_{\mu i}\left(z_{\mu\to i} + F_{\mu i} s_i + \eta_\mu - \omega_{\mu\to i}\right) \\
    &= \sum_\mu F_{\mu i}^2s_i + \sum_\mu F_{\mu i}\left(z_{\mu\to i} + \eta_\mu - \omega_{\mu\to i}\right)
\end{split}
\end{equation}
The first piece concentrates to a deterministic value:
\begin{equation}
    \mathbb{E}\left[\sum_\mu F_{\mu i}^2s_i\right] = \frac{1}{N}\sum_\mu s_i = \alpha s_i
\end{equation}
The second one converges to a Gaussian variable. Let's compute its mean and variance:
\begin{equation}
    \mathbb{E}\left[ \sum_\mu F_{\mu i}\left(z_{\mu\to i} + \eta_\mu - \omega_{\mu\to i}\right) \right] = 
    \sum_\mu \mathbb{E}[F_{\mu i}]\mathbb{E}\left[\sum_{j\neq i} F_{\mu j} s_j + \eta_\mu - \omega_{\mu\to i}\right] = 
    0
\end{equation}

\begin{equation}
\begin{split}
    \mathbb{E}\left[\sum_\mu F_{\mu i}^2\left(z_{\mu\to i} + \eta_\mu - \omega_{\mu\to i}\right)^2 \right] &=\frac{1}{N}\sum_\mu \mathbb{E}[(\eta_\mu)^2] + \mathbb{E}\left[\left(z_{\mu\to i} - \omega_{\mu\to i}\right)^2\right] =\\
    &=\frac{1}{N}\sum_{\mu} \left(\Delta^2 + E\right) =\\
    &=\alpha \left(\Delta^2 + E\right)
\end{split}
\end{equation}

We conclude that the messages $m_i(x_i)$ are sampled from \textit{randomly distributed} Gaussian variables, with mean $R_i$ and variance $\Sigma_i$:
\begin{equation}
R_i = s_i + z \sqrt{\frac{\Delta^2 + E}{\alpha}}\,, \qquad  \Sigma_i = \frac{\Delta^2 + v}{\alpha}
\end{equation}
where $z\sim \mathcal{N}(0,1)$ is a normal Gaussian variable.
We can finally obtain an expression for the order parameters. Calling $\mathcal{D}z$ the normal Gaussian measure:
\begin{equation}
    \mathcal{D}z = \frac{dz}{\sqrt{2\pi}}e^{-\frac{z^2}{2}}
\end{equation}
and $\mathcal{D}m(v,E)$ the Belief measure:
\begin{equation}
    \mathcal{D}m(v,E) = \,\frac{\mathcal{D}z \, dx \,ds\, P^x(x) P^s(s)}{\mathcal{Z}(s,v,E)} \exp{\left\{-\frac{\alpha}{2\left(\Delta^2 + v\right)}\left(x - s - z \sqrt{\frac{\Delta^2 + E}{\alpha}}\right)^2\right\}}
\end{equation}
\begin{figure}[b!]
\centering
\includegraphics[width=0.475\columnwidth]{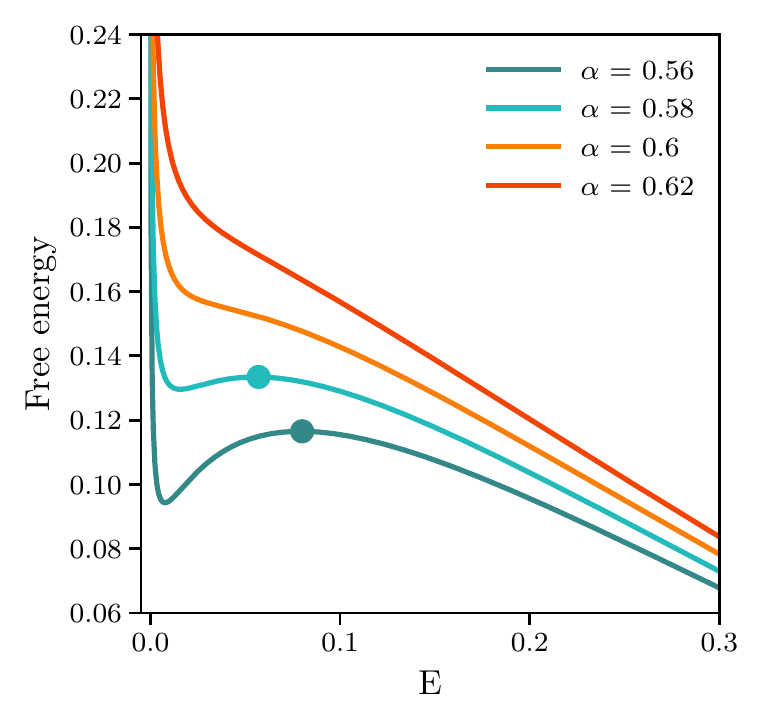}
\hfill
\includegraphics[width=0.475\columnwidth]{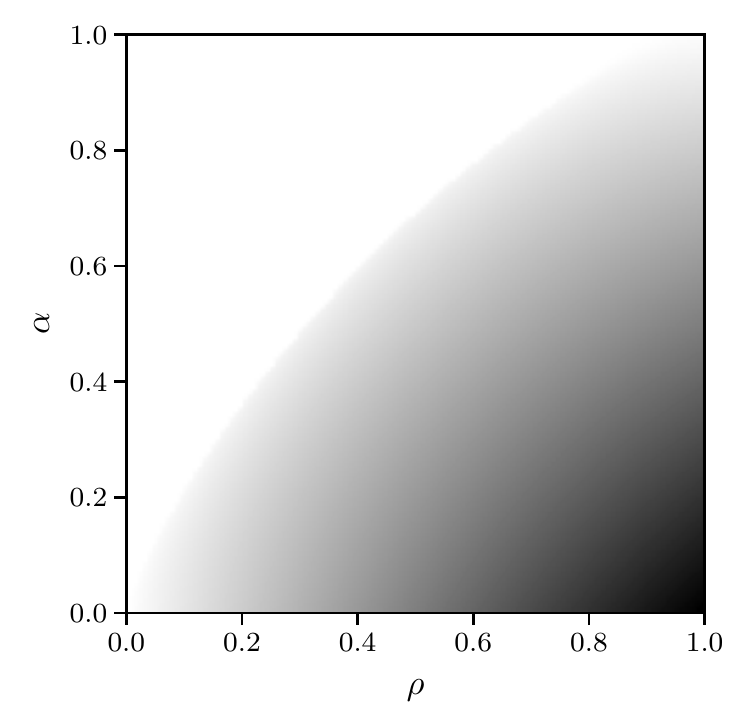}
\hfill
\caption{
Asymptotic performance of noiseless $\Delta\to0$ compressed sensing solved using AMP. On the right: asymptotic error as a function of the sparsity $\rho$ and the sample complexity $\alpha$. White is zero error. On the left: free energy as a function or the error for $\rho = 0.4$ and several values of the sample complexity. The dot is the value where AMP would converge from a random initialisation. This figure is reproduced independently from \cite{Krzakala_2012}
}\label{fig:Phase_diagram}
\end{figure}
By using the definitions of the order parameters we have:
\begin{equation} \label{SE_CS}
\begin{split}
    v &= \int \left[x - \left(\int x\, \mathcal{D}m(v,E)\right)\right]^2\, \mathcal{D}m(v,E)\\
    E &= \rho \,\mathbb{E}[s^2] + \int \left(x^2 - 2xs \right)\, \mathcal{D}m(v,E)\\
\end{split}
\end{equation}

Equation \eqref{SE_CS} can again be turned into an algorithm called State Evolution, which can be used to reveal the expected performace of AMP. In fact, state evolution can be seen as a gradient ascent method in the two-dimensional space $v,E$, trying to reach the maximum of a free-entropy function $\Phi(v,E)$. This free-entropy can be derived from the equations, but, interestingly, it can also be obtained by the replica analysis of the next section. We give an example of this in Figure \ref{fig:Phase_diagram}, where we analysed the noiseless $\Delta\to 0$ limit of compressed sensing. From the free energy plotted as a function of $E$ we can clearly see that for each $\rho$ there is a minimal sample complexity $\alpha_{AMP}(\rho)$ above which the problem is solved exactly by AMP, as the gradient ascent state evolution flows towards $E\to 0$. Notably, this AMP threshold is $\alpha_{AMP}(\rho)>\rho$: the algorithm is suboptimal, it is not able to find the zero error state when $\rho<\alpha<\alpha_{AMP}(\rho)$, even if we know that the problem is  solvable in that regime by exponential algorithms. The free energy tells us why this isn't happening: for values of $\alpha$ that are too low a local maxima appears. This prevents a randomly initialised AMP to reach zero error, generating a \textit{hard} region. This is how $\alpha_{AMP}(\rho)$ is computed

It is possible to derive the State Evolution of AMP for the general model in a very similar way to what shown above.
First, with the same argument as in equation \eqref{E_concentration} we find that $z_{\mu\to i}$ and $\omega_{\mu\to i}$ converge jointly to a Gaussian distribution:
\begin{equation}
    \begin{pmatrix}
    z_{\mu\to i} \\
    \omega_{\mu\to i}
    \end{pmatrix}
    \approx
    \begin{pmatrix}
    z \\
    \omega
    \end{pmatrix}
    \sim \mathcal{N} \left(
    \begin{bmatrix}
    0 \\
    0
    \end{bmatrix},
    \begin{bmatrix}
    \rho \,\mathbb{E}[s^2] & m \\
    m & q
    \end{bmatrix}
    \right)
\end{equation}
with $m$ and $q$ defined as:
\begin{equation}
\begin{split}
    m &= \frac{1}{N}\sum_{i=1}^N s_i\hat x_i \\    
    q &= \frac{1}{N} \sum_{i=1}^N \hat x_i^2 \\
\end{split}
\end{equation}
As before $V_{\mu\to i}$ will concentrate on $v$. 
In the general model the asymptotics of $\sum_\mu A_{\mu \to i}$ is
\begin{equation}
    \sum_\mu\mathbb{E}\left[A_{\mu \to i}\right] = - \frac{\alpha}{P}\sum_{\mu}G_2^\mu(V_{\mu \to i},\omega_{\mu \to i}) \approx \alpha\, \mathbb{E}_{\omega, G_2}\left[ - G_2(v,\omega) \right] = \hat{v}
\end{equation}
where the last expectation is taken over $\omega$ and all sources of randomness inside $G_2$, as for example $y^\mu$ in the perceptron or compressed sensing case. We thus include the dependence on $z$ in $G_1$ and $G_2$, leading to the definition:
\begin{equation} \label{eq:SE_vhat}
    \hat{v} = \alpha \,\mathbb{E}_{\omega, z}\left[ - G_2(v,\omega, z) \right]
\end{equation}
Similarly, we can look at $\sum_\mu B_{\mu \to i}$
\begin{equation}
\begin{split}
    \sum_\mu\mathbb{E}\left[B_{\mu \to i}\right] &= F_{\mu i}\sum_{\mu}G_1^\mu(V_{\mu \to i},\omega_{\mu \to i}, z_{\mu\to i} + F_{\mu i}s_i) \approx\\
    &\approx F_{\mu i}\sum_{\mu}G_1^\mu(V_{\mu \to i},\omega_{\mu \to i}, z_{\mu\to i}) + s_i F_{\mu i}^2 \sum_{\mu}\frac{\partial G_1^\mu}{\partial z}(V_{\mu \to i},\omega_{\mu \to i}, z_{\mu\to i})\\
    &\approx \zeta \sqrt{ \hat{q}} +  s_i \hat{m} \\
\end{split}
\end{equation}
with $\zeta \sim \mathcal{N}(0,1)$ and
\begin{equation} \label{eq:SE_mhat}
\begin{split}
    \hat{m} &= \alpha\,\mathbb{E}_{\omega, z}\left[\frac{\partial G_1}{\partial z}(v,\omega, z)\right]\\ 
    \hat{q} &= \alpha\,\mathbb{E}_{\omega, z}\left[{G_1}^2(v,\omega, z)\right]\\ 
\end{split}
\end{equation}
The messages $m_i(x_i)$ have mean $R_i$ and variance $\Sigma_i$ with:
\begin{equation}
R_i = \zeta \sqrt{\frac{\hat{q}}{\hat{v}^2}} + s_i \frac{\hat{m}}{\hat{v}} \,, \qquad  \Sigma_i = \frac{1}{\hat{v}}
\end{equation}
We can finally close the equations by recalling the measure \eqref{eq:ref_measure}
\begin{equation} \label{eq:SE_m}
\begin{split}
    m &= \mathbb{E}_{s,\zeta}\left[s \int x\, \mathcal{D}m(R,\Sigma)\right]\\
    q &= \mathbb{E}_{s,\zeta}\left[\left(\int x\, \mathcal{D}m(R,\Sigma)\right)^2\right] \\
    v &= \mathbb{E}_{s,\zeta}\left[\int x^2\, \mathcal{D}m(R,\Sigma)\right] - q \\
\end{split}
\end{equation}
So for the general model we would iterate equations \eqref{eq:SE_m}, \eqref{eq:SE_mhat}, \eqref{eq:SE_vhat} until convergence, then the observables would be written as a combinations of $m$, $q$ and $v$.

\section{Replica method for the general model}
It is possible to obtain equivalent expressions for the overlaps derived though state evolution by using the replica method. Recall the partition function of the general model \eqref{proba_general}:

\begin{equation}  
  \mathcal{Z} = \int\text{d}\bx  \prod_{i=1}^NP^\mathbf{x}_i(x_i)\prod_{\mu=1}^P\Psi^\mu \left(\mathbf{F}^\mu \cdot\mathbf{x}\right)
\end{equation}
We will assume here for simplicity that all matrix elements $F^\mu_i$ are chosen independently from a Gaussian distribution with mean zero and variance $1/N$ (in fact all our reasoning applies to any i.i.d. distribution with these first and second moment, provided higher moments are well defined).
Our goal will be to compute the free entropy by using the replica method:

\begin{equation}
    \lim_{N\to\infty} \frac{1}{N}\mathbb{E}[\log{ \mathcal{Z}}] = \lim_{n\to 0} \frac{\lim_{N\to\infty} \frac{1}{N} \mathbb{E}[\mathcal{Z}^n]-1}{n}
\end{equation}
where $\mathbb{E}$ denotes the average with respect to the distribution of $F$.

Let's thus compute the $n$-th moment of the partition function by considering $n$ independent replicas of the system indexed by $a\in\{1,...,n\}$:

\begin{equation}
  \mathbb{E}[\mathcal{Z}^n] = \mathbb{E}\;\left( \int\prod_{a=1}^n\prod_{i=1}^N\left[\text{d}x_i^a     P^\mathbf{x}_i(x_i^a)\right] \int \prod_{a=1}^n\prod_{\mu=1}^P\left[\text{d}z^a_\mu\,\Psi^\mu \left(z_\mu^a\right)\delta(z_\mu^a-\mathbf{F}^\mu \cdot\mathbf{x}^a)]\right]\right)
\end{equation}
were we introduced the $P\times n$ auxiliary variables $z^a_\mu$. The average in this case is over $\mathbf{F}^\mu$. In some cases there is another source of randomness  like the labels in the perceptron or the data in the compressed sensing case. We can account for it by considering it an extra replica with index $a=0$. Now, using the central limit theorem we find that $z_\mu^a$ are joint Gaussian variables with mean and covariance:
\begin{equation}
\begin{split}
    &\mathbb{E}\left[\sum_{i=1}^NF_i^\mu x_i^a\right] = 0\,, \\
    &\mathbb{E}\left[\sum_{i,j}F_i^\mu F_j^\nu x_i^ax_j^b\right] = \frac{\delta_{\mu\nu}}{N} \sum_{i=1}^N x_i^ax_j^b = \delta_{\mu\nu}Q^{ab}\,,\\
\end{split}
\end{equation}
where $Q^{ab}=(1/N)\sum_i x_i^a x_i^b$ is going to be our order parameter. Notice that while the replicas were independent they all became coupled after averaging over the disorder. We can now write:
\begin{equation}
  \mathbb{E}[\mathcal{Z}^n] = \prod_{1\leq a\leq b\leq n}\int\text{d}Q^{ab}\, I_{\mathrm{prior}}(\mathbf{Q}) I_{\mathrm{coupling}}(\mathbf{Q})
\end{equation}
where:
\begin{align}
   I_{\mathrm{prior}}(\mathbf{Q})&= 
   \int\prod_{a=1}^n\prod_{i=1}^N\left[\text{d}x_i^a     P^\mathbf{x}_i(x_i^a)\right]\; \prod_{a\leq b} \delta\left( Q^{ab}-\frac{1}{N}\sum_i x_i^a x_i^b\right)\\
   I_{\mathrm{coupling}}(\mathbf{Q})&=\prod_{\mu=1}^P\left[\int \prod_{a=1}^n\left[\text{d}z^a_\mu\,\Psi^\mu \left(z_\mu^a\right)\right] 
   \; \frac{1}{\sqrt{2\pi\text{det}\mathbf{Q}}}e^{-\frac{1}{2}\sum_{a,b}Q^{-1}_{ab} z_\mu^a z_\mu^b}
   \right]
\end{align}
Notice that the quantity $ I_{\mathrm{coupling}}(\mathbf{Q})$ is a product of $P$ independent integrals. As for $I_{\mathrm{prior}}(\mathbf{Q})$, it can also be written as $N$ independent integrals if one first introduces integral representations of the $\delta$ functions. Neglecting prefactors that are non-exponential in $N$, we get
\begin{align}
   I_{\mathrm{prior}}(\mathbf{Q})&= \int \prod_{a\leq b}  d{\hat{Q}}^{ab} e^{-\frac{N}{2}\sum_{a,b} \hat{Q}^{ab}Q^{ab}} \prod_{i=1}^N\left[
   \int\prod_{a=1}^n\text{d}x^a     P^\mathbf{x}_i(x^a)\; e^{\frac{1}{2}\sum_{a,b} \hat{Q}^{ab} x^a x^b}\right]
\end{align}
in order to keep notations simple, let us assume that $P^\mathbf{x}_i$ is independent of $i$ and $ \Psi^\mu$ is independent of $\mu$ (the results can be easily extended to the more general case). We obtain:
\begin{equation}
  \mathbb{E}[\mathcal{Z}^n] = 
\int\prod_{a\leq b}\left[\text{d}Q^{ab}\text{d}\hat{Q}^{ab}\right]
e^{N\Phi(\mathbf{\hat{Q}},\mathbf{Q} ) }
\label{Zngen}
  \end{equation}
where
\begin{align}
\Phi(\mathbf{\hat{Q}},\mathbf{Q} )&=-\frac{1}{2} \text{Tr}( \mathbf{\hat{Q}}\mathbf{Q})+ \Phi_{\mathrm{prior}}(\mathbf{\hat{Q}})+\alpha \Phi_{\mathrm{coupling}}(\mathbf{Q})
\\
\Phi_{\mathrm{prior}}(\mathbf{\hat{Q}})&= \log\left( \int\prod_{a=1}^n\left[\text{d}x^a     P^\mathbf{x}(x^a)\right]\; e^{\frac{1}{2}\sum_{a,b} \hat{Q}^{ab} x^a x^b}\right)
\\
\Phi_{\mathrm{coupling}}(\mathbf{Q})&=\log\left( 
\int \prod_{a=1}^n\left[\text{d}z^a\,\Psi \left(z^a\right)\right] 
   \; \frac{1}{\sqrt{2\pi\text{det}\mathbf{Q}}}e^{-\frac{1}{2}\sum_{a,b}Q^{-1}_{ab} z^a z^b}\right)
\end{align}
and $\alpha =\frac{P}{N}$. The form (\ref{Zngen}) gives the expression of the average of $Z^n$ for the general model. At large $N$, it can be computed with the saddle point method: one needs to find two $n\times n$ symmetric matrices $\mathbf{\hat{Q}}$ and $\mathbf{Q}$ which are a stationary point of the function $\Phi(\mathbf{\hat{Q}},\mathbf{Q} )$.

In the "replica symmetric Ansatz", one restricts the search to matrices which are symmetric under permutation of the replicas. Then 
\begin{align}
Q_{ab}&= r\delta_{ab} + q(1-\delta_{ab}) \,,\qquad\qquad \hat{Q}_{ab} = -\hat{r}\delta_{ab} + \hat{q}(1-\delta_{ab})
\end{align}
Essentially we traded an optimisation over the entire matrices $\mathbf{Q}$ and $\mathbf{\hat{Q}}$ for one over four scalar quantities. It is a good exercise to check that this general formalism gives back the known replica symmetric expressions for the various applications of the general model described previously. For instance, in the Hopfield model where $x_i$ are Ising spins and $\Psi^\mu(z)=e^{\beta z^2/2}$, one can check that our general formalism gives back the standard replica symmetric formulation of the glassy phase found by Amit et al.

We might of course take much more general ansatz \cite{Parisi79}, this one will be sufficient for the problem at hand.
All the quantities above simplify. Starting with the trace term we have:
\begin{equation}
    \text{Tr}( \mathbf{\hat{Q}}\mathbf{Q}) = -n\hat{r}r + n(n-1)\hat{q}q
\end{equation}
To compute the determinant of $\mathbf{Q}$ we notice that 
\begin{equation}
    \mathbf{Q} = q + (r-q)\delta_{ab}
\end{equation} 
The eigenvalues of $\mathbf{Q}$ are thus $r + (n-1)q$ with multiplicity $1$ and $r - q$ with multiplicity $n-1$, leading to:
\begin{equation}
    \text{det}\mathbf{Q} = (r + (n-1)q)(r - q)^{n-1}
\end{equation}
Noticing that $\mathbf{1}^2 = n\mathbf{1}$, where $\mathbf{1}$ is the matrix with all ones, we get:
\begin{equation}
    \mathbf{Q}^{-1} = -\frac{q}{(r-q)(r + q(n-1))} + \frac{1}{r-q}\delta_{ab}
\end{equation} 
The prior integral becomes
\begin{align}
    \Phi_{\mathrm{prior}}(\mathbf{\hat{Q}})&= \log\left( \int\prod_{a=1}^n\left[\text{d}x^a     P^\mathbf{x}(x^a)\right]\left[\exp{\left\{-\frac{\hat{r}+\hat{q}}{2}\sum_{a=1}^n (x^a)^2 + \frac{\hat{q}}{2}\left(\sum_{a=1}^n x^a\right)^2\right\}}\right]\right)\\
    &= n\log\left( \mathbb{E}_{x,\zeta}\left[\exp{\left\{-\frac{\hat{r}+\hat{q}}{2} \; x^2 + \sqrt{\hat{q}}\; x\zeta\right\}}\right]\right)
\end{align}
where we used the identity $\mathbb{E}_\zeta[\exp(\sqrt{c}\zeta)] = e^\frac{c}{2}$. 
The constraint integral is done similarly. At the saddle point we can take the limit $n\to 0$, giving the free entropy $\Phi$:

\begin{align}
    \Phi &= \frac{\hat{q}q}{2} + \frac{\hat{r}r}{2} + \Phi_{\mathrm{prior}}(\hat{q},\hat{r}) + \alpha\Phi_{\mathrm{coupling}}(q,r)
    \\
    \Phi_{\mathrm{prior}}(\hat{q},\hat{r})&= \mathbb{E}_{\zeta}\log\left( \mathbb{E}_{x}\left[\exp{\left\{-\frac{\hat{r}+\hat{q}}{2} x^2 + \sqrt{\hat{q}}x\zeta\right\}}\right]\right)
    \\
    \Phi_{\mathrm{coupling}}(q,r)&= \mathbb{E}_{\zeta}\log\left( \left[\int\mathrm{d}z\frac{\Psi(z)}{\sqrt{2\pi(r-q)}}\exp{\left\{-\frac{(z+\sqrt{q}\zeta)^2}{2(r-q)}\right\}}\right]\right)
\end{align}
evaluated at the solutions of the following system:
\begin{align}
    q &= -2\frac{\partial \Phi_{\mathrm{prior}}}{\partial \hat{q}}\,, \qquad\qquad \hat{q} = -2\alpha \frac{\partial \Phi_{\mathrm{coupling}}}{\partial q}\\ 
    r &= -2\frac{\partial \Phi_{\mathrm{prior}}}{\partial \hat{r}}\,, \qquad\qquad \hat{r} = -2\alpha \frac{\partial \Phi_{\mathrm{coupling}}}{\partial r}
\end{align}
Including the $a=0$ replica will just change the procedure above slightly. The replica ansatz in this case will be:

\begin{equation}
\mathbf{Q}=\left(\begin{array}{cccc}
r_0 & m & \ldots & m \\
m & r & \ldots & \ldots \\
\cdots & \cdots & \cdots & q \\
m & \cdots & q & r
\end{array}\right)\,,
\qquad\quad
\mathbf{\hat Q}=\left(\begin{array}{cccc}
-\hat r_0 & \hat m & \ldots & \hat m \\
\hat m & -\hat r & \ldots & \ldots \\
\cdots & \cdots & \cdots & \hat q \\
\hat m & \cdots & \hat q & -\hat r
\end{array}\right)
\end{equation}
The auxiliary quantities are:
\begin{equation}
    \text{Tr}( \mathbf{\hat{Q}}\mathbf{Q}) = -\hat{r}_0r_0 -n\hat{m}m -n\hat{r}r + n(n-1)\hat{q}q
\end{equation}
\begin{equation}
    \text{det}\mathbf{Q} = (r - q)^{n-1}\left[r_0(r + (n-1)q) - nm^2\right]
\end{equation}
and for the inverse matrix $\mathbf{Q}^{-1}$:
\begin{equation}
\mathbf{Q}^{-1}=\left(\begin{array}{cccc}
\tilde r_0 & \tilde m & \ldots & \tilde m \\
\tilde m & \tilde r & \ldots & \ldots \\
\cdots & \cdots & \cdots & \tilde q \\
\tilde m & \cdots & \tilde q & \tilde r
\end{array}\right)
\end{equation}
with:
\begin{align}
    \tilde r_0 &= \frac{r+(n-1)q}{r_0(r + (n-1)q) - nm^2} \\
    \tilde m &= -\frac{m}{r_0(r + (n-1)q) - nm^2}\\
    \tilde q &= - \frac{q}{(r-q)(r + q(n-1))} - \frac{m \tilde{m}}{r + q(n-1)}\\
    \tilde r &= \frac{1}{r-q} - \frac{q}{(r-q)(r + q(n-1))} - \frac{m \tilde{m}}{r + q(n-1)}\\
\end{align}
The prior and coupling integrals become:
\begin{align}
    &\Phi_{\mathrm{prior}}(\hat{m},\hat{q},\hat{r})= \mathbb{E}_{\zeta}\log\left( \mathbb{E}_{x,x^*}\left[\exp{\left\{-\frac{\hat{r}_0}{2}(x^*)^2 + \hat{m}x^*x-\frac{\hat{r}+\hat{q}}{2} x^2 + \sqrt{\hat{q}}x\zeta\right\}}\right]\right)
\end{align}
\begin{equation}
\begin{split}
\Phi_{\mathrm{coupling}}(m,q,r)=& \\
\mathbb{E}_{\zeta}\log\left( \int\mathrm{d}z\,\mathrm{d}z^*\right.&\left.\frac{\Psi(z)\Psi(z^*)}{\sqrt{2\pi\,\mathrm{det}{\bf Q}}}\exp{\left\{-\frac{\tilde{r}_0}{2}(z^*)^2 -\tilde{m}z^*z - (\tilde{r} - \tilde{q})\frac{z^2}{2} + \sqrt{\tilde{q}}z\zeta\right\}}\right)
\end{split}    
\end{equation}
Leading to the system:
\begin{align}
    q &= -2\frac{\partial \Phi_{\mathrm{prior}}}{\partial \hat{q}}\,, &\hat{q} = -2\alpha \frac{\partial \Phi_{\mathrm{coupling}}}{\partial q}&\\ 
    r &= -2\frac{\partial \Phi_{\mathrm{prior}}}{\partial \hat{r}}\,, &\hat{r} = -2\alpha \frac{\partial \Phi_{\mathrm{coupling}}}{\partial r}&\\
    m &= \frac{\partial \Phi_{\mathrm{prior}}}{\partial \hat{m}}\,, &\hat{m} = \alpha \frac{\partial \Phi_{\mathrm{coupling}}}{\partial m}&
\end{align}
The order parameters found with this replica approach are simple functions of the ones found in the state evolution section, upon setting $v = r - q$ and $\hat v = \hat r + \hat q$. The equations for $m$, $q$ and $v$ are easily obtained by taking a  derivative, the other three are less immediate, and we just show as an example how to derive the equations in the case without the $a=0$ replica, leaving the rest as an exercise.

The first step is to rewrite the channel integral in a more manageable form. Notice that:
\begin{equation}
    \int\mathrm{d}z\frac{\Psi(z)}{\sqrt{2\pi V}}e^{-\frac{(z+\sqrt{q}\zeta)^2}{2V}} = \frac{1}{2\pi}\iint\mathrm{d}z\mathrm{d}k \frac{\tilde{\Psi}(k)}{\sqrt{2\pi V}}e^{-\frac{z^2}{2V} + z(ik-\sqrt{q}\zeta/V)-q \zeta^2/(2V)}
\end{equation}
Upon integrating on $z$ we find that the expression above is $G_0(r-q,\sqrt{q}\zeta)$. The coupling integral thus becomes:
\begin{equation}
        \Phi_{\mathrm{coupling}}(q,r) = \mathbb{E}_{\zeta}\log G_0(r-q,\sqrt{q}\zeta)
\end{equation}
We are ready to take the derivative with respect to $r$ and $q$.
Recall the identity:
\begin{equation}
    \frac{d G_0(V,\omega)}{d V} = \frac{1}{2}\frac{d^2 G_0^\mu(V,\omega)}{d\omega^2}
\end{equation}
We thus have 
\begin{equation}
    \frac{\partial \Phi_{\mathrm{prior}}}{\partial q} = \mathbb{E}_{\zeta}\left[ -\frac{1}{G_0 (V,\omega)}\frac{d G_0(V,\omega)}{d V}\bigg{\rvert_{\substack{V = r-q\\\omega = \sqrt{q}\zeta}}} + \frac{\zeta}{2\sqrt{q}}\frac{1}{G_0 (V,\omega)}\frac{d G_0(V,\omega)}{d \omega}\bigg{\rvert_{\substack{V = r-q\\\omega = \sqrt{q}\zeta}}} \right]
\end{equation}
To proceed we need Stein's lemma. Let $\zeta\sim\mathcal{N}(0,1)$ be a normal Gaussian variable, then:
\begin{equation}
    \mathbb{E}_{\zeta}\left[f(\zeta)\zeta\right] = \mathbb{E}_{\zeta}\left[ f'(\zeta) \right]
\end{equation}
hence:
\begin{equation}
    \mathbb{E}_{\zeta}\left[ \zeta G_1(r-q,\sqrt{q}\zeta) \right] = \sqrt{q} \,\mathbb{E}_{\zeta}\left[ G_2(r-q,\sqrt{q}\zeta) \right]
\end{equation}
In conclusion we have:
\begin{equation}
    \hat{q} = \alpha \mathbb{E}_{\zeta}\left[ {G_1}^2(r-q,\sqrt{q}\zeta) \right]
\end{equation}
Similarly we have:
\begin{equation}
    \hat{r} = \alpha \mathbb{E}_{\zeta}\left[ -{G_2}(r-q,\sqrt{q}\zeta) \right] - \hat{q}
\end{equation}
As anticipated we obtain exactly the state evolution equations by setting after a redefinition $v = r-q$ and $\hat{v} = \hat{r} + \hat{q}$.

\section{Outlook: dictionary learning and matrix factorisation}
We conclude this exposition with a brief discussion on dictionary learning.
Suppose you have a $M\times P$ matrix of samples $y^{\mu s}$ generated by:

\begin{equation}
    y^{\mu s} = {\mathbf{F}^*}^\mu \cdot {\mathbf{s}^*}^s + \eta^{\mu s}
\end{equation}
where $\eta^{\mu s}$ is i.i.d. noise, $\mathbf{F}^* \sim P^{\mathbf{F}}$, $\mathbf{s}^* \sim P^{\mathbf{s}}$. We wish to recover both $\mathbf{F}$ \textit{and} $\mathbf{x}$. 
We can think of this problem as a generalisation of compressed sensing in which we also try to recover the measurement matrix,t thus the case in which the apparatus is unknown. In the noiseless case this is also a matrix factorisation problem. We could try to use the same approach as before and write the posterior distribution $P(\mathbf{s},\mathbf{F}|y)$

\begin{equation}
    P(\mathbf{s},\mathbf{F}|y) = \prod_{\mu, i} dP^F(F_i^\mu) \prod_{s,i} dP^X(x_i^s) e^{-\frac{1}{2\Delta^2} \sum_{\mu, s}(y^{\mu s} - \frac{1}{\sqrt{N}}\sum_i F_i^\mu x_i^s)^2}
\end{equation}
If $N$ is finite, while $M, P \to \infty$ then this problem is extensively studied, see \cite{Lesieur_2017} and references therein. Indeed, if $y^{\mu s}$ is a planted signal, my best estimate for the underlying signal and the apparatus will be given by the expectation values $\langle x^s_i\rangle$ and $\langle F^{\mu}_i\rangle$.  If however also $N$ is infinite, while
\begin{equation}
    \alpha = M/N\,, \qquad \psi = P/N
\end{equation}
are finite the naive AMP approach fails \cite{Kabashima_2016}. A number of approaches have been attempted \cite{Maillard_2022, Barbier_2022}, but at the present time extensive rank matrix factorisation remains an open problem.

\ack
We thank Mario di Luca for his assistance on the figures and Paula Anette Mürmann for carefully reading the draft.
These are notes from the lecture of Marc Mézard given at the summer school "Statistical Physics \& Machine Learning" that took place in Les Houches School of Physics in France from 4th to 29th July 2022. The school was organized by Florent Krzakala and Lenka Zdeborová from EPFL.




\bibliographystyle{iopart-num}
\bibliography{biblio}

\end{document}